\documentclass[twoside]{sfm}

\input{sf.def}

\begin{document}

\def\llm{{\sc LLmodels}}
\def\atl{{\sc ATLAS9}}
\def\aatl{{\sc ATLAS12}}
\def\starsp{{\sc STARSP}}
\def\aur{$\Theta$~Aur}
\def\logg{\log g}
\def\tauros{\tau_{\rm Ross}}
\def\kms{km\,s$^{-1}$}
\def\bz{$\langle B_{\rm z} \rangle$}
\def\degr{^\circ}
\def\kk{$\kappa$ Cnc}
\def\aaps{A\&AS}
\def\aap{A\&A}
\def\apjs{ApJS}
\def\apj{ApJ}
\def\Rh{\rule{20.0pt}{0.0pt}}
\def\Rv{\rule[-0.1in]{0.0pt}{20.0pt}}
\def\Sum{N_{\rm tot}}
\def\rmxaa{Rev. Mexicana Astron. Astrofis.}
\def\mnras{MNRAS}
\def\actaa{Acta Astron.}
\newcommand{\Tef}{T$_{\rm eff}$~}
\newcommand{\Vt}{$V_t$}
\newcommand{\CC}{$^{12}$C/$^{13}$C~}
\newcommand{\CDC}{$^{12}$C/$^{13}$C~}
\newcommand{\hangpar}{\noindent\hangindent.5in}%

\pagebreak

\thispagestyle{titlehead}

\setcounter{section}{0}
\setcounter{figure}{0}
\setcounter{table}{0}

\markboth{Cowley}{Concluding remarks}
\titl{Putting the A-stars into context: concluding remarks}
{Cowley, C.$^1$}
{$^1$University of Michigan, Ann Arbor, MI, USA, email: {\tt cowley@umich.edu} 
}
\renewcommand{\thefootnote}{\arabic{footnote}}

\baselineskip 12pt

\section{Preamble}
I want to thank the SOC for inviting me to give 
the final presentation, summarizing our conference.
At the meeting I tried to mention all of the talks,
and do my best to pronounce the speaker's names.  
I appreciate the help I got.
There is no need to review
the presentations in these written final remarks.  The reader may simply turn to
the appropriate authoritative presentations themselves.  Instead, we shall make a number of 
comments on aspects of CP star abundances, and 
our understanding of them from the prospectus of
half a century's efforts in this field.

\section{Two areas}

Before turning specifically to abundances and
spectroscopy, I should like to express admiration
for developments on the theoretical and observational
fronts in two areas: pulsation, and magnetic/chemical
mapping.  Developments in these areas are formidable,
amazing, and hold great promise.

\section{The A-star puzzle}

The chemically peculiar A- and related stars have 
been a puzzle for more than a century.  In these
remarks, we shall mostly discuss certain aspects
of the chemistry of those CP stars known as HgMn
or mercury-manganese stars.
Among the CP stars, they
have the simplest atmospheric structures, with
minimal complications due to magnetic fields or
convection.  At the same time, their abundances show
the most extreme deviations from the solar pattern.

It was reasonable consider the largest anomalies
first. In attempting to understand any set of 
observational results, it is wise to begin with
the extreme cases.  Once they are understood, 
hopefully, the intermediate ones will fall into 
line.

The diffusion theory introduced by Georges Michaud
(\cite{GM70}) and the Vauclairs (cf. Vauclair, et
al. \cite{VVSM}) was immediately seen to be able to
account for abundance anomalies.  It explained the
overall patterns, which have the light, abundant
elements depleted and the heavier, more rare ones in
excess. It was soon shown to to account for even the
most extreme anomalies, ca. 6 dex, in a time as
short as a million years. Already by the mid 1970's
there were papers written to account for isotopic
anomalies in helium and mercury.

One might conclude, from the lack of recent efforts
on the extreme and most bizarre (e.g. isotopic)
anomalies that this area was well understood.  This
is not the case.

\section{Credit where it is due}

Before proceding with specifics, it is well to give
credit to a few abundance workers upon whose efforts
we draw. The field is dominated by papers of Adelman
(see \cite{SJA03}, \cite{APWLD} and papers
referenced), Ryabchikova, and their coworkers
(\cite{RZA}, and \cite{TAR13}). Kudos are due to
Dworetsky and Keith Smith for a large body of work
notably with the IUE spectra (\cite{KCSMMD},
\cite{KCS96}, \cite{MMDPP}). As a result of these
truly monumental efforts, we now have a clear idea
of the range of abundance variations, from element
to element, and from star to star.  This work built
on the qualitative but nevertheless excellent work
of Bidelman \cite{WPB}.

We also note the relatively recent discovery by
Castelli and Hubrig (\cite{CasHub}) of the rare
isotope of calcium, $^{48}$Ca.  It is not well
understood how this nuclide exists in nature at all
(cf. Clayton \cite{DD}), much less why it has been
identified in stellar spectra. It is observed in
both magnetic and HgMn stars, and appears to be the
dominant isotope in HR 7143.

On the theoretical side, Georges Michaud has earned
and received wide recognition for his work.  We are
pleased to be able to add our own accolades, to him,
the Vauclairs, Georges Alecian, and their coworkers
(cf. Alecian, et al. \cite{40Y}). If their enormous
efforts have not yet completely solved the problem
of CP star abundances, they have provided the solid
background for a solution.

Very fine observational and theoretical work has
been carried out by workers whom we have no space
to mention individually.  We offer our apologies.

\section{Confrontation with observations}

The early theoretical work relative to HgMn stars
gave predictions of many elements (cf. Michaud, et
al. \cite{MCVV}). Attention was focused on the
atmospheres and outer stellar envelopes.  

Since the late 1990's diffusion has been integrated
into evolutionary calculations for complete stellar
models.  Ironically, the level of accuracy acceptable
for the most of the models, was not achievable
for the outermost atmospheric layers--with
temperatures below about 200,000K (Turcotte \&
Richard \cite{TR}). We currently await results with
a realistic treatment of atmospheres and microscopic
physics for these regions(cf. Stift \& Alecian \cite{SA}, and
Alecian, et al. \cite{ASD}).

We predict that the new results, when they come,
will not be {\em qualitatively} different from the old
ones.  There will still be a need to slow the 
diffusion mechanism down, and prevent abundance 
excesses much larger than those observed.  If
this prediction is correct, the ad hoc turbulent
diffusion coefficient $D_T$ will still be needed.
 
In the modern, extensive work on Am/Fm stars, the
region with temperatures below 200,000K is assumed
to be thoroughly mixed.  Chemical separation takes
place in deeper layers, where simpler
approximations, e.g.  LTE, and a coarser opacity
grid, are acceptable.  So far, the results have been
judged promising, though significant problems
remain.  However, the AmFm stars do not have extreme
anomalies of heavier elements.  

Observers know, on the other hand, that among the
non- (or mildly-) magnetic HgMn stars, abundance
ratios can vary significantly from one star to
another.  Abundance differences occur even between
stars with similar temperatures. For example, on the
iron peak, the star 53 Tau has a large (2 dex)
manganese excess. Indeed, the Cr-Mn-Fe triplet
itself shows a non-nuclear, odd-Z anomalous
abundance pattern (see \cite{KCS96}, and papers
referenced therein).

However, the Hg abundance in 53 Tau is
virtually solar, nor are there the P or Ga
excesses often seen in HgMn stars. Mu Lep, with
nearly the same temperature and Mn excess has
overabundant P and Ga, and is overabundant in
mercury by some 5 dex.

Why, while diffusion processes were creating the Mn
excess in 53 Tau, did they not cause other excesses
typical of HgMn stars, such as Hg, P, or Ga?  

A useful case concerns a star whose
unusual abundances were unknown a decade ago, HD
65949.  This star may have the most extreme mercury
overabundance known.  The elements Pt, Os, and Re
are in excess by 5 to 6 dex.  The expected
depletions of the light elements O and C are not
seen, though N is deficient by 2 or more dex.  This
star has about the same effective temperature as
$\kappa$ Cnc or HR 7143 but it's Mn abundance is less than that of those stars by  1.5 to 1.8 dex
(cf. Cowley, et at. \cite{CHPQB}).

Surely a complete understanding of the chemistry of
the HgMn and related stars will explain not only the
extreme abundances, but also why, in the same stars,
other groups of elements have relatively
unfractionated patterns.  The sophisticated
examination of radiative support of Hg by Proffitt
et al. \cite{Proff}, like that of Michaud, Reeves,
and Charland \cite{MRC} before it, treated only Hg
or its isotopes. This was also the case for the
study of Sapar et al. \cite{Sapar}.

It is essential to check the wider consequences of 
any model.  For example, we mentioned that even the
small fraction of the heavy isotope $^{48}$Ca present 
in nature is problematical.  But it is easy to set up 
neutron-rich, nuclear quasi-equilibrium 
conditions that would lead to the meteoritic 
$^{48}$Ca/$^{40}$Ca ratio.  The problem is that this
model would lead to other isotopic abundances that
are not observed in nature.  For our stars, it is
thus essential to ask what might be happening to 
other elements while the atmospheric conditions are
such that subtle isotopic fractionation can be taking
place.

\section{A token search for order\label{sec:order}}
Let us consider two simple scenarios for the
development of abundance anomalies in CP stars.  In
both cases, the initial abundances are solar, but in
the first case the anomalies develop very rapidly
and reach an equilibrium so that intermediate
patterns are rarely seen.  In this case, differences
in abundance patterns would be seen in stars with
different temperatures.  Those with the same
temperatures should have similar abundances.

In the second case, we assume abundance anomalies
develop over time.  In this case, stars with similar
temperatures should fall into sequences, with the
youngest (or those with the least chemical
differentiation) at one end and the oldest at the
other.  One should be able to put the spectra in
order, much as astronomy students sort stars into
the temperature sequence by the appearance of their
spectra.
\begin{table}[t]
\centering
\caption{Abundance data [El]\label{tab:abs}}
\begin{tabular}{llllllll}\hline
Star    &  T  & [P] & [S]  &  [Cr]  & [Mn]& [Fe] & [Hg] \\ \hline
46 Aql  &13000&+1.5 &$-$1.8  &$-$1.5 &+0.8&+0.6  &+3.8 \\
HD 65949&13100&+1.5 &$-$1.0  & +0.5  &+0.6&+0.5  &+6.3 \\
\kk     &13225&+1.8 &$-$1.0  & +0.1  &+2.4&$-$0.2&+4.9\\
HR 7361 &13300&+2.0 &$-$1.0  & +0.2  &+2.5&+0.1  &+4.4 \\ \hline
\end{tabular}
\end{table}


Table ~\ref{tab:abs} gives logarithmic abundance
differences from the sun for four HgMn or related
stars. All values are from the second spectra (first
ions), from \cite{RZA}, \cite{CHPQB}, and
\cite{TAK}. The effective temperatures are similar
though in Castelli's \cite{FC} recent analysis 46
Aql has $T_e = 12560$ K. Numerical values are in the
usual bracket notation, and rounded to one decimal.
\begin{table}
\centering
\caption{Abundance Sorts} \label{tab:absor} 
\begin{tabular}{llllll}
\hline   
by P    & by S   &  by Cr  &  by Mn  &  by Fe   & by Hg  \\ \hline
46 Aql  &46 Aql  &46 Aql   &HD 65949 &\kk       &46 Aql \\
HD 65949&\kk     &\kk      &46 Aql   &HR 7361   &HR 7361 \\ 
\kk     &HR 7361 &HR 7361  &\kk      &HR 65949  &\kk     \\
HR 7361 &HD 65949&HD 65949 &HR 7361  &46 Aql    &HD 65949 \\ \hline
\end{tabular}
\end{table}


In Table ~\ref{tab:absor}, the stars are sorted by
increasing abundances of the elements in the column
headings.  So, for example, 46 Aql has the least
enhanced phosphorus (P), and HR 7361, the most.
The stars $\kappa$ Cnc and HR 7361 might
be called classical HgMn stars, and 
their abundances are similar.  The other
two stars are distinct from this pair, and from one
another.

It is impossible to explain the the varying orders
of Table ~\ref{tab:absor} in terms of a simple model
where all stars begin with a solar abundance, and
rapidly differentiate to a uniform abundance
pattern.  Naively, one might think from Table
~\ref{tab:absor} that 46 Aql was the least
chemically differentiated--the least mature--of the
four stars.  But it is the richest in Fe. Moreover,
its S and Cr show significant depletions.

Both 46 Aql and HD 65949 are highly differentiated,
but in quite distinct ways.  It is difficult to see
how, given time, one abundance pattern would evolve
into the other.  If time is the significant factor,
we can ask which elements begin to show their
anomalies first.  We see no clue to a temporal order
in the tables presented here. A more complex
scenario is needed.

\section{Other peculiar stars}
\subsection{Cobalt stars}
One hears almost nothing of the cobalt stars, whose
extreme members are found in the magnetic CP
sequence. Some of these stars have Co/Fe ratios
greater than unity: HR 4950, HR 1094, HD 200311 (cf.
Nishimura \cite{Nish}).  The triple Fe-Co-Ni shows
an odd-Z anomaly reminiscent of the Cr-Mn-Fe pattern
seen in some manganese stars.  These anomalies 
surely arise by chemical separation processes.
Why are they so
rare?  What unusual processes might cause the enhancement
of cobalt rather than manganese?
\subsection{Lambda Boo stars}

There was little beyond the brief mention by Landstreet
\cite{JL1} about the $\lambda$ Boo stars at our
symposium.  Was that because our main tool for
understanding CP stars can't help us with these stars?
For the typical diffusion pattern is a depletion of the
abundant lighter elements with an excess of the rarer,
heavier ones.  Overall, this fits most CP stars rather
well.  But the $\lambda$ Boo stars have the opposite
pattern--normal lighter elements with depletions of the
heavier ones.

Still, diffusion must be relevant!  As has been 
emphasized many times, the mechanism is fundamental
and must work.  No one has suggested the
atmosphere/envelope of the $\lambda$ Boo stars is
convective!  What has been suggested is that
grain-depleted gas has somehow been added to their
surfaces.  This hypothesis predicts a depletion of
refractory elements, but not volatile ones.

This $\lambda$ Boo hypothesis has a serious flaw.
The volatile element zinc (Zn) is depleted in those 
stars where its abundance has been determined.  This
has been known for some time, but is rarely discussed
(cf. Heiter \cite{Heiter}).
Could diffusion account for this?  We note here that
the Zn abundance is one of the most highly variable
among the CP stars.  In the HgMn star $\chi$ Lup,
it was found to be depleted by some 4 orders of
magnitude (Leckrone et al. \cite{Leck1}).
But Zn is generally found to be 
significantly underabundant in most HgMn stars
(Smith \cite{KSZn}).  If
it fell on a stellar photosphere, would it diffuse
quickly away?

\section{Exogenous processes}
Forty years ago, the notion of material falling on a
main-sequence star was regarded as virtually
impossible.  Today, we know it happens. 
We have seen spectacular images of comets or 
asteroids being devoured by the sun (cf.
\cite{youtube}).  In most cases, events like
that cited are of disruptions in the low corona,
but Sekanina \cite{Sek} discusses the plausibility
of direct impacts.
We also know
that debris falls on white dwarfs.  Indeed,
atmospheric abundances in certain white dwarfs may
give an indication of the nature of the
circumstellar material (G\"{a}nsicke, et al.
\cite{Gans},  Jura \& Xu \cite{Jura}).

There is gathering evidence that our sun itself may
bear the complementary pattern to that shown by the
metal-rich white dwarfs (Mel\'{e}ndez
\cite{JMelen}). The most accurate solar/stellar
abundances indicate the sun is depleted in
refractory elements.  This, essentially, is a
subdued example of the abundance pattern already
noted for the $\lambda$ Boo stars--refractory
elements depleted, volatile elements (apart from Zn)
normal.

We are aware of debris disks about young and even
mature stars.  Could the puzzling star spot on 
$\alpha$ And have been caused by infalling material?
(cf. Korhonen \cite{Heidi})

Mainstream theoretical efforts to understand CP 
abundances have been largely limited to 
endogenous processes, those
taking place in the stellar atmospheres,
interiors, or in winds.
Accretion, mass transfer in binaries,
and debris infall have not been included in 
detailed calculations.  
Some of these newly recognized phenomena might
work well in synergism with chemical separation
mechanisms.  

Modern calculations show time-dependent abundance
oscillations in stellar interiors (cf. Th\'{e}ado,
et al. \cite{TALV}).  Could they introduce sufficient
disorder to frustrate a time sequencing of 
atmospheric abundance
patterns we sought in Section ~\ref{sec:order}? The
question is moot, but not a justification to eschew
exogenous processes that we know are plausible.

I thank Gautier Mathys for his comments on an early
version of this paper.


\begin{thebibliography}{}
\bibitem{SJA03}
{Adelman, S. J., Adelman, A. S. \& Pintado, O. I.}
2003, A\&A, 397, 267.

\bibitem{APWLD}
{Adelman, S. J., Proffitt, C. R., Wahlgren, G. M.,
Leckrone, D. S. \& Dolk, L.} 2004, ApJS, 155, 179.

\bibitem{ASD}
{Alecian, G., Stift, M. J. \& Dorfi, E. A.} 2011,
MNRAS, 418, 986.

\bibitem{40Y}
{Alecian, G., Richard, O. \& Vauclair, S. (ed)} 2005,
40 years of atomic diffusion: meeting in honor of
Georges Michaud, Ch\^{a}teau de Mons, France (EAS
Pub. Ser. v. 17).

\bibitem{WPB}
{Bidelman, W. P.} 1966, in Abundance Determinations in Stellar Spectra, IAU Symp. 26, ed. H. Hubenet (London: Academic Press), p. 229.

\bibitem{FC}
{Castelli, F.} 2013, http://wwwuser.oat.ts.astro.it/castelli/

\bibitem{CasHub}
{Castelli, F. \& Hubrig, S.} 2004, A\&A, 421, L1.

\bibitem{DD}
{Clayton, D.} 2004, Handbook of isotopes in the cosmos: [V.1] Hydrogen to Gallium (see p. 196).

\bibitem{CHPQB}
{Cowley, C. R., Hubrig, S., Palmeri, P., et al.} 2010, MNRAS, 405, 1271.


\bibitem{MMDPP}
{Dworetsky, M. M., Persaud, J. L. \& Patel, K.}
2008, MNRAS, 385, 1523.

\bibitem{Gans}
{G\"{a}nsicke, Koester, D., Girven, J., et al.} 2012,
MNRAS, 424,333.

\bibitem{Heiter}
{Heiter, U.} 2002, A\&A, 381, 959.

\bibitem{Jura}
{Jura, M. \& Xu, S.} 2013, AJ, 145, 30.

\bibitem{Heidi}
{Korhonen, H.} 2013, this volume.

\bibitem{JL1}
{Landstreet, J. D. } 2013, this volume.

\bibitem{Leck1}
{Leckrone, D. S., Proffitt, C. R., Wahlgren, G. M., et al.} 1999, AJ, 117, 1454.

\bibitem{JMelen}
{Melendez, J.} 2013, arXiv:1307.5274

\bibitem{GM70}
{Michaud, G.} 1970, ApJ, 160, 641.

\bibitem{MRC}
{Michaud, G., Reeves, H. \& Charland, Y.} 1974, A\&A,
37, 313.

\bibitem{MCVV}
{Michaud, G., Charland, Y., Vauclair, S. \& Vauclair, G.} 1976, ApJ, 210, 477.

\bibitem{Nish}
{Nishimura, M., Sadakane, K., Kato, K., Takeda, Y. \& Mathys, G.} 2004, A\&A, 420, 673.

\bibitem{Proff}
{Proffitt, C. H., Brage, T., Leckrone, D. S., et al.}
1999, ApJ, 512, 942.

\bibitem{TAR13}
{Ryabchikova, T. A.} 2013, this volume.

\bibitem{RZA}
{Ryabchikova, T. A., Zakharova, L. A. \& Adelman, S. J.} 1996, MNRAS, 283, 1115.

\bibitem{Sapar}
{Sapar, A., Aret, A., Poolam\"{a}e, R. \& Sapar, L.}
2008, Contrib. Astron. Obs. Skalnat\'{e} Pleso, 38,
273.

\bibitem{Sek}
{Sekanina, Z.} 2002, ApJ, 566, 577.

\bibitem{KSZn}
{Smith, K. C.} 1994, A\&A, 291, 521.

\bibitem{KCS96}
{Smith, K. C.} 1996, Astrophys. Sp. Sci., 237, 77. 

\bibitem{KCSMMD}
{Smith, K. C. \& Dworetsky, M. M.} 1993, A\&A, 274, 335.

\bibitem{TAK}
{Sadakane, K., Takada-Hidai, M., Takada, Y., et al.}
2001, PASJ, 53, 1223

\bibitem{SA}
{Stift, M. J. \& Alecian, G.} 2012, MNRAS, 425, 2715.

\bibitem{TALV}
{Th\'{e}ado, S., Alecian, G., LeBlanc, F. \&
Vauclair, S.} 2012, A\&A, 546, 100.

\bibitem{TR}
{Turcotte, S. \& Richard, O.} 2008, in
40 years of atomic diffusion: meeting in honor of
Georges Michaud, Ch\^{a}teau de Mons, France (EAS
Pub. Ser. v. 17, ed. G. Alecian, O. Richard \&
S. Vauclair) 

\bibitem{youtube}
{youtube} 2013, http://www.youtube.com/watch?v=GwusUex\_tRA.

\bibitem{VVSM}
{Vauclair, S., Vauclair, G., Schatzman, E.,
\& Michaud, G.} 1978, ApJ, 223, 567.

\end{thebibliography}
\end{document}